\documentclass[aps,prl,nobalancelastpage,twocolumn,superscriptaddress,nolongbibliography,10pt]{revtex4-2}

\usepackage{color,amsthm,amsmath,amsxtra,amsfonts,dsfont,graphicx,bm,amssymb}
\usepackage[normalem]{ulem}
\usepackage[colorlinks=true,linkcolor=blue, citecolor=blue, urlcolor=blue, bookmarks]{hyperref}
\usepackage{centernot}
\usepackage[dvipsnames]{xcolor}
\usepackage{tikz}
\usepackage{braket}
\usepackage{subfigure}
\usepackage{bm}
\usepackage{bigints}
\usepackage[normalem]{ulem}
\usepackage{bbold}
\usepackage{mathrsfs}

\newcommand{\be}{\begin{equation}}
\newcommand{\ee}{\end{equation}}
\newcommand{\bea}{\begin{eqnarray}}
\newcommand{\eea}{\end{eqnarray}}
\newcommand{\bse}{\begin{subequations}}
\newcommand{\ese}{\end{subequations}}

\newcommand{\prlsection}[1]{{\em {#1}.---~}}

\begin{document}

\title{A bound on approximating non-Markovian dynamics by tensor networks in the time domain}

\author{Ilya Vilkoviskiy}
\affiliation{Department of Theoretical Physics, University of Geneva, Quai Ernest-Ansermet 30, 1205 Geneva, Switzerland}
\author{Dmitry A. Abanin}
\affiliation{Department of Theoretical Physics, University of Geneva, Quai Ernest-Ansermet 30, 1205 Geneva, Switzerland}
\affiliation{Google Research, Mountain View, CA, USA}

\begin{abstract}
Spin-boson (SB) model plays a central role in studies of dissipative quantum dynamics, both due its conceptual importance and relevance to a number of physical systems. Here we provide rigorous bounds of the computational complexity of the SB model for the physically relevant case of a zero temperature Ohmic bath. We start with the description of the bosonic bath via its Feynman-Vernon influence functional (IF), which is a tensor on the space of spin's trajectories. By expanding the kernel of the IF functional via a sum of decaying exponentials, we obtain an analytical approximation of the continuous bath by a finite number of damped bosonic modes. We bound the error induced by restricting bosonic Hilbert spaces to a finite-dimensional subspace with small boson numbers, which yields an analytical form of a matrix-product state (MPS) representation of the IF. We show that the MPS bond dimension $D$ scales polynomially in the error on physical observables $\epsilon$, as well as in the evolution time $T$, $D\propto T^4/\epsilon^2$. This bound indicates that the spin-boson model can be efficiently simulated using polynomial in time computational resources. 

\end{abstract}
\maketitle
\prlsection{Introduction}
Every quantum system, irrespective of its nature, interacts with an environment. While in some cases environment dynamics can be described in the Markovian approximation, there is a broad class of problems where this is not sufficient, and non-Markovian effects are essential. Examples of such phenomena range from non-equilibrium transport in quantum dots~\cite{madsen2011observation,mccutcheon2011general} to micromechanical resonators~\cite{groeblacher2015observation} and light harvesting complexes~\cite{huelga2013vibrations,potovcnik2018studying}. 

An archetypical model of non-Markovian quantum dynamics is the spin-boson (SB) model, describing a spin-$1/2$ coupled to a harmonic bosonic bath. As discussed in the seminal work of Leggett {\it et al.}~\cite{Leggett:1987zz}, this model exhibits a rich variety of dynamical regimes, including a dissipative phase transition. Despite a variety of analytical results obtained in various limits, the exact description of system's dynamics in the strong coupling limit, which is needed to model realistic physical systems, remained an outstanding challenge. 

To address this challenge, a variety of numerical methods has been developed. In one direction, Chin {\it et al.}~\cite{chin2010exact} used a chain mapping of a continuous bosonic bath, afterwards modeled by tDMRG techniques. Another family of approaches is based on truncating the Feynman-Vernon influence functional (IF), or a related object, augmented reduced density tensor (ADT). While an early quasiadiabatic propagator
path integral (QUAPI) algorithm~\cite{Makri1995TensorPF} had exponential scaling with the memory time, more recently efficient schemes based on tensor-networks compression have been introduced~\cite{strathearn2018efficient,jorgensen2019exploiting,luchnikov2020machine,bose2022multisite,cygorek2023sublinear,link2023open}, see also Ref.~\cite{Makri} for the improved QUAPI algorithm. In another direction, it was argued~\cite{tamascelli2018nonperturbative,somoza2019dissipation,mascherpa2020optimized} that the IF can be approximated by a bath of a finite number of auxiliary bosons, or by a finitely many  hierarchical equations of motion (HEOM), which also leads to a drastic boost of computational efficiency~\cite{xu2022taming}. It was concluded that the system can be simulated efficiently. We note in passing that related ideas were developed in the context of fermionic quantum impurity models~\cite{dorda2014auxiliary,thoenniss2023efficient,ng2023real,thoenniss2023nonequilibrium}, and for interacting environments~\cite{PhysRevX.11.021040,ye2021constructing}.

Despite these remarkable developments, no theoretical bounds on the complexity of simulating the spin-boson model exist, and the goal of this paper is to fill this gap. Focusing on the case of zero temperature ohmic bath, we introduce a number of (damped) auxiliary bosons approximating the original bath. As a main result, we show that the Feynman-Vernon influence functional, as well as system's evolution are efficiently simulated by the auxiliary modes. We provide an analytical expression for the matrix-product state (MPS) approximation of the influence functional, and prove that the bond dimension scales polynomially with time:
\begin{equation} \label{BondD}
   D \sim \frac{(\omega_cT)^4}{\epsilon^2},
\end{equation} 
where $\epsilon$ is the error for the physical observables, i.e. the density matrix of the spin at time $T$.

Finding a representation in terms of auxiliary of modes is equivalent to approximating certain correlation functions in terms of a finite sum of decaying exponents. Each exponent leads to a single bosonic mode, which then can be truncated to a finite dimensional subspace. The decomposition of an arbitrary function in terms of decaying exponents is a well known problem in theory of signal processing and mathematical physics, which can be efficiently solved numerically with the help of Prony's method~\cite{MarpleDigitalSpectralAnalysis}. For our purposes we use another approach by Beylikin and Monzon ~\cite{beylkin2010approximation}, which allows for analytical estimates, see Supplementary Online Material (SOM)~\cite{SM} for the review.
%
%

\prlsection{Influence functional description of the spin-boson model} The main object of our interest will be SB model consisting of a spin impurity coupled to a bosonic bath at zero temperature. Its Hamiltonian reads:
\begin{equation}    H=H_{\text{b}}+H_{\text{int}}+H_{\text{s}},
\end{equation}
with
\begin{equation}
    H_{\text{b}}=\sum\limits_{k} \omega_k a^{\dagger}_ka_k,
\end{equation}
\begin{equation}
    H_{\text{int}}=c_k(a_k^{\dagger}+a_k)\sigma_z.
\end{equation}
Our method is suitable for arbitrary spin Hamiltonian $H_{\text{s}}(t)$, which is not specified. 

We consider discrete time dynamics, obtained by the Trotterization of the original model. Evolution operator over a small time step $\Delta t$ is approximately given by:
\begin{equation}
U_{\text{total}}=e^{i \Delta t (H_{\text{b}}+H_{\text{int}})}e^{i\Delta t H_{\text{s}}(t_k)}+O(\Delta t^2).
\end{equation}
As we are interested in continuous dynamics, we assume $\Delta t\sim \epsilon_1=\frac{\epsilon}{T}$. It is useful to introduce a shorthand notation for the exponentials in the above expression:
\begin{equation}
e^{i \Delta t (H_{\text{b}}+H_{\text{int}})}=U_{\text{b}}(a,a^{\dagger}|\sigma_z),
\end{equation}  
\begin{equation}
e^{i\Delta t H_{\text{s}}(t_k)}=U^{(k)}.  
\end{equation}
For a fixed evolution time $T=N\Delta t$, we can integrate out the bosonic degrees of freedom defining a discrete analog of the Feynman-Vernon influence functional (IF) for a fixed trajectory of a spin:
\begin{equation}
\mathcal{I}_{s_N,\bar{s}_N,\dots,s_1,\bar{s}_1}=\ _{\text{bath}} \! \langle 0|\prod\limits_{k=1}^{N-1} U_{\text{b},\bar{s}_k}^{\dagger}\rho\prod\limits_{k=1}^{N-1} U_{\text{b},s_k}|0\rangle_{\text{bath}},
\end{equation}
where $U_{\text{b},s_k}\overset{\text{def}}{=}\langle s_k|U_{\text{b}}(a,a^{\dagger}|\sigma_z)|s_k\rangle$. Then, we rewrite the time-evolved density matrix in terms of IF:
\begin{multline}\label{IM_dynamics}
\tilde{\rho}_{s_N,\bar{s}_N}=\\=\!\sum\limits_{s_k,\bar{s}_k=1,2}\mathcal{I}_{s_N,\bar{s}_N,\dots,s_1,\bar{s}_1}\prod\limits_{k=1}^{N-1}\left( U^{(k)}_{s_k,s_{k+1}}U^{(k)\, \dagger}_{\bar{s}_{k+1},\bar{s}_k}\right) \rho_{s_1,\bar{s}_1},
\end{multline}
The IF of the spin trajectories $\mathcal{I}_{s_N,\bar{s}_N,\dots,s_1,\bar{s}_1}\equiv \mathcal{I}_{\{\boldsymbol{s},\boldsymbol{\bar{s}}\}}$ can be computed explicitly ~\cite{Makri1995TensorPF}
\begin{equation}\label{IM functional}
\mathcal{I}_{\{\boldsymbol{s},\boldsymbol{\bar{s}}\}}=e^{-\sum\limits_{i\le j}\left[(s_i-\bar{s}_i)(s_j\eta_{i,j}-\eta^{\star}_{i,j}\bar{s}_j) \right]},
\end{equation}
where $\eta_{i,j}=\kappa_{i,j}+i\phi_{i,j}$ is a known function: 
\begin{equation}\label{eta}
    \eta_{i,j}=\eta_{i-j}=4\int\limits_{0}^{\infty}d\omega J(\omega)\frac{\sin^2(\frac{\omega \Delta t}{2})}{\omega^2}e^{i\omega t_{i-j}}\,,\quad i>j
\end{equation}
\begin{equation}\label{eta2}
    \eta_{i,i}=2\int\limits_{0}^{\infty}d\omega J(\omega)\frac{\sin^2(\frac{\omega \Delta t}{2})}{\omega^2}\,,
\end{equation}
with $t_{i-j}=(i-j)\Delta t$. We note in passing that our analysis is applicable to the case of a finite $\Delta t$, but we will mostly be interested in the limit $\Delta t\to 0$.
Below we will concentrate on the case of an Ohmic bath, for which the spectral density takes the following form:
\begin{equation}
    J(\omega)=\alpha\omega e^{-\frac{\omega}{\omega_c}}.
\end{equation}

\prlsection{MPS representation for the IF via auxiliary bosons}\label{Section ZMKMP}
To represent the IF as an MPS we will use the approach of Ref.~\cite{zaletel2015time}, which we briefly review below. We introduce the action functional $\mathcal{I}=e^{\mathcal{S}}$:
\begin{equation}
\mathcal{S}=-\sum\limits_{i\le j}\left[(s_i-\bar{s}_i)(s_j\eta_{i,j}-\eta^{\star}_{i,j}\bar{s}_j) \right],
\end{equation}
where function $\eta_{i,j}$ decays polynomially, as $\frac{1}{(\Delta t(i-j))^2}$ for the Ohmic case.  
Let us fix site $i$ and split the action $\mathcal{S}$ to a sum of three different terms:
\begin{equation}
    \mathcal{S}=\mathcal{S}_i^{L}+\mathcal{S}_i^{R}+\sum\limits_{m=1}^K h^L_{i,m}h^R_{i,m},
\end{equation}
where $\{\mathcal{S}_i^{L},h^L_{i,m}\}$ depend only on the variables to the left of the site $i$, and $\{\mathcal{S}_i^{R},h^R_{i,m}\}$ depend on the variables to the right of the site $i$. 

{Suppose the existence of the following relation} ~\cite{zaletel2015time}:
\begin{equation}\label{Recurrence For S}
\begin{pmatrix}
\mathcal{S}_{i-1}^R\\
\boldsymbol{h}^R_{i-1}\\
\mathbb{1}_{i-1}
\end{pmatrix}=\begin{pmatrix} \mathbb{1} & \boldsymbol{C}^i & {D}^i\\
0 & \boldsymbol{A}^i & \boldsymbol{B}^i \\
0& 0& \mathbb{1}
\end{pmatrix}\begin{pmatrix}
\mathcal{S}_i^R\\
\boldsymbol{h}^R_{i}\\
\mathbb{1}_i
\end{pmatrix}.
\end{equation}
Here $\boldsymbol{h}_i^R$ is a row of numbers $h_{i,m}^R$, $\boldsymbol{A}^i$ is an $K \times K$ matrix, $\boldsymbol{C}^{i}$ and $\boldsymbol{B}^i$ are $1\times K$ column and $K\times 1$ row respectively, and $D^i$ is a scalar.
First, the above relation yields an MPS representation for the action $\mathcal{S}$. We start from the site $N+1$, for which $\mathcal{S}^R_{N+1}=0$. Then we move to the left, using the recurrence relation \eqref{Recurrence For S}
\begin{equation}\label{rec1}
    \boldsymbol{h}_{i-1}^R=\boldsymbol{A}^{i}\boldsymbol{h}_{i}^R+\boldsymbol{B}^{i},
\end{equation}
\begin{equation} \label{rec2}
    \mathcal{S}_{i-1}^R=\mathcal{S}_{i}^R+\boldsymbol{C}^{i}\boldsymbol{h}^R_{i}+D^{i}.
\end{equation}
$\mathcal{S}^R_0$ is nothing but the total action $\mathcal{S}$. 

To represent an exponent of the total action $e^{\mathcal{S}}=e^{\mathcal{S}_0^R}$ in terms of MPS
we introduce $n$ bosonic modes: 
\begin{gather}
\boldsymbol{a}=\{a_1,\dots a_K\}^{\scriptscriptstyle{T}}, \\ \boldsymbol{a}^{\dagger}=\{a^{\dagger}_1,\dots a^{\dagger}_K\},
\end{gather}
with the standard commutation relations:
\begin{equation}
    [a_i,a^{\dagger}_j]=\delta_{i,j}.
\end{equation}
We also introduce coherent states: \begin{equation}
|\boldsymbol{h}_i^R\rangle=e^{\sum\limits_{m=1}^K h^R_{i,m}a^{\dagger}_m}|0\rangle.
\end{equation}
The operations \eqref{rec1},\eqref{rec2} can be written in terms of the operator
\begin{equation}\label{BosonicOper}
M^i=e^{D^i} e^{\left(\boldsymbol{a}^{\dagger} \cdot \boldsymbol{B}^i \right)}e^{\left(\boldsymbol{a}^{\dagger} \cdot\log(\boldsymbol{A}^i) \boldsymbol{a}\right)}e^{\left(\boldsymbol{C}^{i} \cdot \boldsymbol{a}\right)},
\end{equation}
{which implicitly depends on $s_i,\bar s_i$.}
Indeed, this operator can be equivalently defined by its action on the coherent states:
\begin{equation}
M^i|\boldsymbol{h}\rangle=e^{D^i+\left(\boldsymbol{C}^i\cdot \boldsymbol{h}\right)}|\boldsymbol{B}^i+\boldsymbol{A}^i\boldsymbol{h}\rangle.
\end{equation}
Applying it to the state $e^{\mathcal{S}^R_i}|\boldsymbol{h}_i^R\rangle$, we get:
\begin{equation}
M^i e^{\mathcal{S}^R_i}|\boldsymbol{h}_i^R\rangle=e^{\mathcal{S}^R_{i-1}}|\boldsymbol{h}_{i-1}^R\rangle.
\end{equation}
Finally, the MPS representation of the IF is provided by the formula
\begin{equation}\label{MPS_representation}
I_{\boldsymbol{s},\boldsymbol{\bar{s}}}=\langle 0|M_{s_1,\bar{s}_1}^1 ,\dots M_{s_N,\bar{s}_N}^N |0\rangle.
\end{equation}
Formally, this yields an MPS representation of IF. Next, we will show that the IF of an Ohmic bath can be well-approximated by choosing a finite $K$. We will further truncate operator ${M_{s_i,\bar{s}_i}^i}$ to a finite dimensional boson subspace which will give rise to an MPS with a finite bond dimension. Due to the rapid decay of matrix elements with the growing number of excited bosons, this will introduce a controlled error.


\prlsection{Exponentially decaying interaction}\label{sec dec}
Applying this approach to the IF \eqref{IM functional}, we introduce two species of bosonic operators $\boldsymbol{a},\bar{\boldsymbol{a}}$, as well as the second set of $\boldsymbol{A}^i,\boldsymbol{B}^i,\boldsymbol{C}^i$ matrices: $\bar{\boldsymbol{A}}^i,\bar{\boldsymbol{B}^i},\bar{\boldsymbol{C}^i}$, corresponding to forward and backward variables of the IF.

For our purposes it is sufficient to assume that the matrix $\boldsymbol{A}^i$ does not depend neither on $i$ nor on $s$ variables. We also assume it to be diagonalizable.
\begin{equation}
    \boldsymbol{A}=\text{Diag}[e^{-\Omega_1\Delta t},\dots ,e^{-\Omega_K \Delta t}],
\end{equation}
\begin{equation}
    \bar{\boldsymbol{A}}=\text{Diag}[e^{-\Omega^{\star}_1\Delta t},\dots ,e^{-\Omega^{\star}_K \Delta t}],
\end{equation}
with $\Omega_k=\gamma_k+i\omega_k$, $\gamma_k>0$. Futhermore we choose $\boldsymbol{B},\boldsymbol{C}$ to be linear in $s$:
\begin{gather}\label{BCD_parameters}
    \boldsymbol{C}^i=(s_i-\bar{s}_i)\{\lambda_1\Delta t,\dots ,\lambda_K\Delta t\},\\
   \bar{\boldsymbol{C}}^i=-(s_i-\bar{s}_i)\{\lambda^{\star}_1\Delta t,\dots ,\lambda^{\star}_K\Delta t\},\\
\boldsymbol{B}^i=s_i\{\lambda_1\Delta t,\dots ,\lambda_K\Delta t\}^{\scriptscriptstyle{T}},\\
   \bar{\boldsymbol{B}}^i=\bar{s}_i\{\lambda^{\star}_1\Delta t,\dots ,\lambda^{\star}_K\Delta t\}^{\scriptscriptstyle{T}},\\
   D^i=\eta_{i,i}(s_i-\bar{s}_i)^2.
\end{gather}
This will lead to the following ansatz for the IF:
\begin{equation}
    \tilde{I}_{\{\boldsymbol{s},\boldsymbol{\bar{s}}\}}=e^{-\Delta t^2\sum\limits_{0\le i\le j\le N}(s_i-\bar{s}_i)(s_j \tilde{\eta}_{i-j}-(\tilde{\eta})^{\star}_{i-j}\bar{s}_j)},
\end{equation}
with 
\begin{equation}\label{teta}
    \tilde{\eta}_{i-j}=\sum\limits_{k=1}^K\lambda_k^2 e^{-(i-j)\Delta t(\Omega_k) }.
\end{equation}
Each term in the expansion \eqref{teta} gives rise to two bosonic modes, yielding an infinite-dimensional MPS approximation of the IF. 

We note that  for a real $\lambda=|\lambda|$, exactly the same MPS could be generated by an auxiliary quantum channel, see SOM~\cite{SM}. For a complex $\lambda$ the corresponding IF doesn't describe any physical bath. However, our methods to estimate the errors for the observables do not require $\lambda$ to be real and positive, therefore, we keep it arbitrary.

Truncation of the infinite-dimensional MPS \eqref{MPS_representation} to a finite-dimensional one involves two types of errors. First, errors occur due to the truncation of the bosonic modes to a finite dimensional subspace. Second type of error appears due to the inaccuracy of approximating function $\eta_{i-j}$ by the finite sum \eqref{teta}. Before discussing the approximation by a finite sum of exponentials, let us briefly explain the intuition behind the estimate of the errors of the first kind, see SOM~\cite{SM} for a rigorous analysis.

Let us change perspective for a moment and consider the dynamics of bosons in the "environment" of the spin mode. The evolution of bosons is a competition between the driving force $\lambda_k s_i a_k^{\dagger}+\lambda_k^{\star} s_i \bar{a}^{\dagger}_k$ which creates the bosons with rate $\lambda$ and the overall decay $-\gamma_k\left(a_k^{\dagger}a_k+\bar{a}_k^{\dagger}\bar{a}_k\right)$ which damps the wave function component with $n$ bosons with rate $\gamma n$. As a result, processes that involve significantly more than $\frac{|\lambda_k|^2}{\gamma_k^2}$ bosons in the system are strongly suppressed. In fact, {below we derive expressions for $\lambda_k,\gamma_k$} and prove (see SOM~\cite{SM}) that the amplitude to have $n$ bosons is suppressed as $4\nu_{\star}^{\frac{n}{2}}$, with some $\nu_{\star}<1$. This in turn explains why the states with high number of excitations could be neglected.
We conclude that the problem of approximating of the IF \eqref{IM functional} in terms of an MPS is equivalent to expanding function $\eta_{i-j}$ \eqref{eta} in terms of a sum of exponentials \eqref{teta}.

\prlsection{MPS representation for the bosonic bath}\label{Section expansion for the bosonic Bath}
To approximate a power-law function $\eta$ by a finite sum of exponentials, we follow the approach of Ref. ~\cite{beylkin2010approximation}. Let us start with an integral representation in the continuous time limit:
\begin{gather} \label{eta0}
\eta^0(t)=\alpha\int\limits_{0}^{\infty}\omega e^{-\frac{\omega}{\omega_c}} e^{i \omega t} d\omega,\\
\eta^0_{i,j}=\eta^{0}\left(\Delta t(i-j)\right)
\end{gather}
with $t=\Delta t(i-j)$. In SOM~\cite{SM}, by performing a change of variables followed by discretization, we show that this integral is well-approximated by by a sum: 
\begin{equation}
\eta^0(t)\to\alpha\chi \sum\limits_{k=-\infty}^{\infty}  \omega_k^2 e^{-\frac{\omega_k}{\omega_c}}e^{i\omega_k t}, 
\end{equation}
where points $\omega_k$ are situated on a contour in the complex plane, $\omega_k=\omega_ce^{k\chi+\frac{i\pi}{4}}$. This sum may be restricted to a finite number of terms because of the fast decay of the tails. We  prove in SOM~\cite{SM}  that $\eta^0(t)$ may be approximated by a sum of $\sim \log^2(\frac{\omega_cT}{\alpha \epsilon})$ exponentials of the form:
\begin{equation} \label{Exponential sum}
\eta^0(t)= \sum \limits_{k=-N_{\epsilon_1}}^{M_{\epsilon_1}} \lambda_k^2 e^{-\Omega_k t}+\delta \eta(t),
\end{equation}
with 
\begin{equation}\label{lambda_final}
    \lambda_k^2=i\alpha \omega_c^2\chi e^{-\frac{1+i}{\sqrt{2}}e^{(k+1/2)\chi} }e^{2(k+1/2)\chi},
\end{equation}
\begin{equation}\label{Omega_final}
    \Omega_k=\omega_ce^{(k+1/2)\chi}\frac{1-i}{\sqrt{2}}.
\end{equation}
Thanks to the fact that the function is smooth enough, the discretization step $\chi$ grows logarithmically with epsilon $\chi\sim\frac{1}{\log\left(\frac{{\alpha}\omega_cT}{\epsilon}\right)}$.
The discrepancy $\delta \eta(t)$ is given by:
\begin{equation}
\int\limits_{t=0}^T|\delta \eta(t)|<\epsilon_1=\frac{\epsilon}{T}.
\end{equation}
The total number of modes $K=M_{
\epsilon}+N_{\epsilon}$ scales as:
\begin{equation}\label{Number of modes K}
K\sim \log^2\left(\frac{\alpha\omega_cT}{ \epsilon}\right),   
\end{equation}
see SOM~\cite{SM} for details.

For finite, but small $\Delta t$ the IF can be approximated in an MPS form \eqref{MPS_representation}:
\begin{equation}\label{MPSfinal}
M_{s_i,\bar{s}_i}=e^{-\Lambda(s_i-\bar{s}_i)^2}\prod_{k=-N_{\epsilon}}^{M_{\epsilon}}e^{\mathcal{M}_{s_i,\bar{s}_i}^k\Delta t},
\end{equation}
\begin{multline}
\mathcal{M}^k_{s_i,\bar{s}_i}=\\ =-\Omega_k \left(a^{\dagger}_k+\frac{(s_i-\bar{s}_i)}{\sqrt{2}}\frac{\lambda_k}{\Omega_k}\right)\left(a_k-\sqrt{2}s_i\frac{\lambda_k}{\Omega_k}\right)-\\-\Omega_k^{\star}\left(\bar{a}_k^{\dagger}-\frac{(s_i-\bar{s}_i)}{\sqrt{2}}\frac{\lambda_k^{\star}}{\Omega_k^{\star}}\right)\left(\bar{a}_k-\sqrt{2}\bar{s}_i\frac{\lambda_k^{\star}}{\Omega_k^{\star}}\right).
\end{multline}
\begin{equation}\label{Lambda_Final}
    \Lambda=\eta_{0,0}+\sum\limits_{k=-N_{\epsilon}}^{M_{\epsilon}}\text{Re}\left(\frac{\lambda_k^2}{\Omega_k}\right)\Delta t.
\end{equation}
Thus, operator $M_{s,\bar{s}}$ is a product of a bosonic MPS and an operator $e^{-\Lambda(s_i-\bar{s}_i)^2}$ acting on a single site. The value of $\Lambda$ may be estimated via Euler–Maclaurin formula, and it is important to note that it remains positive and of order $\epsilon$~\footnote{Note that, in contrast to the formula \eqref{BosonicOper}, we collect the product of exponents into a single exponent. By the Baker–Campbell–Hausdorff formula, these two representations are equivalent up to a redefinition of parameters. The latter redefinition is negligible in $\Delta t\to 0$ limit, moreover the current formula provide a better approximation for the kernel \eqref{eta},\eqref{eta2}.}.

\prlsection{Bound on the error for spin dynamics} \label{Section Errors} In order to estimate the effects of the inaccuracy in \eqref{Exponential sum}, let us examine the dynamics provided by the MPS \eqref{MPSfinal}-\eqref{Lambda_Final}. We fix spin dynamics, specified by unitary operators $U^{(i)}$ acting on the $2$ dimensional space. The system dynamics may be computed in terms of a product of operators acting on both bosonic and spin degrees of freedom:
\begin{equation}\label{Full_Evolution}
\mathcal{U}^{(i)}_{s_{i+1},\bar{s}_{i+1}|s_i,\bar{s}_i}= M_{s_i,\bar{s}_i} \otimes \left( U^{(i)}_{s_{i+1},s_{i}}\otimes(U^{(i)})^{\star}_{\bar{s}_{i+1},\bar{s}_{i}}\right).
\end{equation}
In particular, density matrix at time $T=N\Delta t$ reads:
\begin{equation}
\rho_{s_N,\bar{s}_N}(T)=_{\text{bosons}}\!\!\langle 0|\left(\prod\limits_{i=1}^N \mathcal{U}^{(i)}\right) |0\rangle_{\text{bosons}}\ \rho_{s_1,\bar{s}_1}.
\end{equation}
An important property of the operator $\mathcal{U}^{(i)}$ is that it does not increase the norm of the vectors. Indeed, let us  note that the eigenvalues of quadratic operators $e^{\mathcal{M}^k_{s_i,\bar{s}_i}}$ are less than $1$. Consequently, the eigenvalues of $M_{s,\bar{s}}^{\dagger}M_{s,\bar{s}}$ are also less than $1$. One can conclude that any bounded vector $|v\rangle$ propagating in time remains bounded:
\begin{equation}\label{Ubounded}
\Big|\prod\limits_{i=n_1}^{n_2} \mathcal{U}^{(i)} |v\rangle \Big|^2\le \langle v|v\rangle.
\end{equation}

Now we are in a position to estimate the error induced by the inaccuracy of approximation of the quadratic kernel $\eta_{i,j}=\eta(i-j)$ for the IF of the form \eqref{IM functional}.
Suppose that we have found an MPS representation for a kernel $\eta^{\epsilon_1}$ such that it is close to the actual kernel in $L_1$ norm:
\begin{equation}
\sum\limits_{j=0}^N|\eta(j)-\eta^{\epsilon_1}(j)|=|\delta\eta|_{L_1}\le\frac{\epsilon_1}{4}=\frac{\epsilon}{4T}.
\end{equation}
The first order error is given by the two point correlator:
\begin{equation}
\delta\rho_T=\sum\limits_{i<j}\langle (s_i-\bar{s}_i)(s_j\delta\eta_{i,j}-\delta\eta^{\star}_{i,j}\bar{s}_j)\rangle,
\end{equation}
where we define the correlators as:
\begin{multline}
\langle\prod\limits_{i\in I} s_{i}\prod\limits_{j\in J}\bar{s}_j\rangle\overset{\text{def}}{=}\sum\limits_{s_a,\bar{s}_a}\prod\limits_{i\in I} s_{i}\prod\limits_{j\in J}\bar{s}_j\times \\ \times\prod\limits_{k=1}^{N-1}\mathcal{U}^{(k)}_{s_{k+1},\bar{s}_{k+1}|s_{k},\bar{s}_k} \rho_{s_1,\bar{s}_1}.
\end{multline}
Property \eqref{Ubounded} guarantees that each correlator is bounded by 1, and so the total error is bounded as:
\begin{equation}
    |\delta\rho_T|\le 4T|\delta\eta|_{L_1}\le\epsilon.
\end{equation}
Similar analysis, for a slightly more general case, was provided in ~\cite{mascherpa2017open}.

To sum up, there are two sources of the errors. One type of errors is due to an inaccuracy in approximation of the function $\eta_{i,j}$ \eqref{Exponential sum}. This error is related to the number of bosonic modes $K$ \eqref{Number of modes K}. Another source of errors stems from truncation of the bosonic modes to a finite dimensional subspace. We show in SOM~\cite{SM} that due to the decay of bosonic modes, the bosonic wave function decays with the number of bosonic excitations $n$ as $\nu_{\star}^{n}$. This allows us to restrict the total number of bosonic excitations as $n_{\star}\sim \frac{\log(\frac{\left(\omega_cT\right)^2}{\epsilon})}{\log(\nu_{\star}^{-1})}$. Combining these two results, after some combinatorics, we arrive at the main result \eqref{BondD}.

\prlsection{Concluding remarks}
In this paper, we develop a non-perturbative approach to the spin-boson model, by introducing an analytical method to approximate a zero temperature ohmic bosonic bath by a number of damped oscillators, with a decay rate \eqref{Omega_final} that controls the memory of the corresponding mode, and the coupling strength \eqref{Lambda_Final}. For a fixed error $\epsilon$, the required number of bosonic modes scales polylogarithmically in $\epsilon$ and evolution time $T$. This explicit construction enables a homogeneous MPS approximation for the Feynman-Vernon functional. We prove that the bond dimension scales at most polynomially with $\epsilon$ and $T$ \eqref{BondD}. Previous numerical results of~\cite{strathearn2018efficient} also suggest the polynomial scaling of bond dimension $D\sim T^{q}$, with non universal $q\in [1,2]$. Note that our MPS description of Feynmam-Vernon influence functional (IF) provides an accurate description of the system's dynamics independently of local spin dynamics and coupling strength; thus, it is natural to expect that bond dimension required for a specific choice of spin's dynamics may exhibit a better scaling.
Apart from theoretical interest, our construction may have a practical applications: in particular, it can be used as a starting point for numerical calculations, as our analytic MPS can be further compressed using SVD decomposition.

Although we proved the bounds for the errors, there are general mathematical constraints on the influence functional. One way to formulate the restriction is to say that the spin dynamics \eqref{IM_dynamics} should provide a CPTP map ~\cite{wilde_2013}, for any unitaries $U^{(k)}$. It seems that our MPS approximation does not in general satisfy this requirement, but it has a sufficiently small deviation from it. It is still unclear for us whether it is possible to provide a physical MPS approximation with the same bond dimension. 

An interesting future direction is to provide estimates for IF of a finite-temperature bath with a generic spectral functions, including cases of sub and super-ohmic baths. For instance, our analysis may be applied to the case of super-ohmic bath, yielding a more favorable scaling of bond dimension compared ot the Ohmic case. However, it remains unclear whether this improvement is sufficient to expect sub-polynomial scaling of bond dimension. We postpone this question to future works. Finally, let us note that our analysis can be extended to a wider class of influence functionals, including non-Gaussian one. This can be achieved by including spin dependence of $\boldsymbol{A}$ matrices in \eqref{BosonicOper}. 

{\bf Acknowledgements.} We thank A. Lerose, J. Motruk, M. Sonner and J. Thoenniss for insightful discussions, collaboration on related topics, and comments on the manuscript, and J. Keeling for useful comments. Support by the European Research Council (ERC)
under the European Union’s Horizon 2020 research and innovation program (grant agreement No.~864597) and by the Swiss National Science Foundation is gratefully acknowledged.
\bibliography{SpinBosonBound}
\appendix
\begin{center}
	\textbf{\large Supplementary Online Material}
\end{center}
Here we provide the details concerning the truncation of the infinite dimensional MPS matrices appearing in the main text 
and prove the corresponding bounds for the bond dimension.\\
\twocolumngrid
\section{Discretization of the integral}
In this section we prove that the function
\begin{equation}
\eta^0(t)=\alpha\int\limits_{0}^{\infty}\omega e^{-\frac{\omega}{\omega_c}} e^{i \omega t} d\omega,
\end{equation}
defined on an interval $[0,T]$, may be represented as a sum of $\sim \log^2(\frac{\alpha \omega_c T}{\epsilon})$  exponential terms, with $L_1$ error bounded by $\epsilon_1=\frac{\epsilon}{T}$. To do this, we mostly follow the ideas of ~\cite{beylkin2010approximation}.

It is instructive to rotate the contour, introducing a phase $\omega \to i\omega e^{-i\phi}$
\begin{equation}
\int\limits_{0}^{\infty}\omega e^{-\frac{\omega}{\omega_c}} e^{i \omega t} d\omega=-\int\limits_{0}^{\infty}e^{-2 i \phi}\omega e^{-\frac{\omega}{\omega_c}(\omega_c t+i)e^{-i\phi}}d\omega.
\end{equation}
Let us then introduce the change of variables: $\omega=\omega_c e^{x}$:
\begin{equation}\label{eta_int}
  \eta^0(t)=-\alpha \omega_c^2\int\limits_{-\infty}^{\infty}e^{-e^{x-i\phi}(\omega_ct +i)}e^{2(x-i\phi)}dx.
\end{equation}
It is useful to denote the integrand of \eqref{eta_int} as $f_t(x)$:
\begin{equation}
    \eta^0(t) \overset{\text{def}}{=}\int\limits_{-\infty}^{\infty}f_t(x) dx.
\end{equation}
The idea is to discretize the integral
\begin{equation}
    \int\limits_{-\infty}^{\infty} f_t(x) dx\to\chi\sum\limits_{k=-\infty}^{\infty}f_t(k\chi).
\end{equation}
Which implies for the $\eta^0$:
\begin{equation}\label{ft}
\eta^0(t)=\sum\limits_k f_t(k\chi)+\delta\eta(t)    
\end{equation}
To estimate the discrepancy $\delta\eta(t)$ we first estimate the difference between the integral and the discrete sum with the help of Poisson re-summation formula:
\begin{equation}
    \chi\sum\limits_{k=-\infty}^{\infty}f_t(x+\chi k)=    \sum\limits_{n=-\infty}^{\infty}\hat{f}_t(\frac{k}{\chi})e^{-\frac{2\pi i k x}{\chi}},
\end{equation}
where $\hat{f}_t(p)$ is the Fourier image of $f_t(x)$:
\begin{equation}
    \hat{f}_t(p)=\int\limits_{-\infty}^{\infty}f_t(x)e^{-2\pi i p x}dx.
\end{equation}
Such that we have:
\begin{equation}
 \Big|\chi\sum\limits_{k=-\infty}^{\infty}f(x+ \chi k)-\int\limits_{-\infty}^{\infty}f_t(x)dx\Big|\le \sum\limits_{k\ne 0}|\hat{f}_t(\frac{k}{\chi})|.
\end{equation}
Fourier transform of our function of interest is equal to:
\begin{equation}
    \hat{f}_t(p)=\alpha \omega_c^2\Gamma(2-2\pi i p) \left(\tau\omega_c\right)^{2\pi i p-2},
\end{equation}
with $\tau=\tau'+i \tau''=(t+\frac{i}{\omega_c})e^{-i\phi}$. We then have:
\begin{multline}
    |\hat{f}_t(p)|\le  \alpha |\Gamma(2-2\pi i p)||\tau|^{-2}e^{-2\pi p \ \text{arg}(\omega_c\tau) }\le \\ \le \alpha |\Gamma(2-2\pi i p)||\tau|^{-2}e^{\frac{\pi^2}{2} p},
\end{multline}
where in the last inequality we put $\phi =\frac{\pi}{4}$, which leads to $|\text{arg}(\omega_c\tau)|<\frac{\pi}{4}$.\\
The gamma function may be estimated as follows:
\begin{equation}
    \Gamma(z)=e^{i\theta z}\int\limits_{0}^{\infty}e^{-\omega e^{i\theta}}\omega^{z-1}d\omega.
    \end{equation}
Consequently, we have:
\begin{equation}
|\Gamma(2-i y)|\le e^{-y \theta}\int\limits_{0}^{\infty}e^{-\omega \cos(\theta)}\omega d\omega=\cos^{-2}(\theta)e^{-y \theta},
\end{equation}
for $\theta<\frac{\pi}{2}$.\\
If we put $\theta=\frac{\pi}{2}-\delta$, we will get:
\begin{equation}
    \Gamma(2-2\pi i p)\le e^{-2\pi (\frac{\pi}{2}-\delta)p}\sin^{-2}(\delta) , \quad \text{for any}\  \delta<\frac{\pi}{2}.
\end{equation}
We thus may restrict the sum $\sum\limits_{k\ne 0}|\hat{f}_t(\frac{k}{\chi})|$ as:
\begin{equation}
\sum\limits_{k\ne 0}|\hat{f}_t(\frac{k}{\chi})|\le 2(|\tau| \sin(\delta))^{-2}\frac{\alpha}{e^{\frac{2\pi}{\chi}(\frac{\pi}{4}-\delta)}-1}.
\end{equation}
Choosing $\chi=\frac{\frac{\pi^2}{2}-2\pi \delta}{\log(\frac{\alpha \omega_c}{\epsilon_1 \sin^2(\delta)})}$, we finally restrict the difference:
\begin{equation}
\Big|\chi\sum\limits_{k=-\infty}^{\infty}f_t(x+ \chi k)-\int\limits_{-\infty}^{\infty}f_t(x)dx\Big|\le \frac{\epsilon_1 \omega_c}{\left(\omega_ct\right)^2+1},
\end{equation}
with {$\epsilon_1=\frac{\epsilon}{T}$}.

In principle it is possible to optimise over $\delta$, but it is not sufficient for our estimate, we simply put $\delta=\frac{\pi}{16}$. This fixes the discretization step $\chi$:
\begin{equation}\label{chi}
\chi=\frac{\pi^2/{8}}{\log\left(\frac{4\alpha\omega_c T}{\epsilon \sin^2(\frac{\pi}{16})}\right)}.
\end{equation}
Now, one can note that the tails in the infinite sum are also decaying, such that one may restrict the summation as follows:
\begin{equation}\label{Beylikin-Monson}
    \Big|\chi\sum\limits_{k=-\infty}^{\infty}f_t(x+\chi k)-\sum\limits_{k=-N_{\epsilon}}^{M_{\epsilon}}f_t(x+ \chi k)\Big|_{L_1}\le \epsilon_1,
\end{equation}
with
\begin{gather}\label{N}
    N_{\epsilon}= \frac{\log\left(\frac{\alpha\omega_c}{\epsilon_1}\right)}{\chi}\sim\log^2\left(\frac{\alpha\omega_c}{\epsilon_1}\right) ,\\ \label{M}
    M_{\epsilon}=\frac{\log\left(\sqrt{2}\log(\frac{\alpha\omega_c }{\epsilon_1})\right)}{\chi}\sim \log\left(\frac{\alpha \omega_c}{\epsilon_1}\right)  \log\left(\log(\frac{\alpha \omega_c}{\epsilon_1})\right).
\end{gather}
Indeed, first we may majorate the absolute value of the sum, by the sum of the absolute values:
\begin{equation}
\Big|\sum\limits_{k=-\infty}^{-N_{\epsilon}}\chi f_t(x+\chi k)\Big|<\sum\limits_{k=-\infty}^{-N_{\epsilon}}\alpha \omega_c^2\chi e^{-e^{k\chi}\tau^{\prime}\omega_c+2k\chi},
\end{equation}
\begin{equation}
\Big|\sum\limits_{k=M_{\epsilon}}^{\infty}\chi f_t(x+\chi k)\Big|<\sum\limits^{\infty}_{k=M_{\epsilon}}\alpha\omega_c^2\chi e^{-e^{k\chi}\tau^{\prime}\omega_c+2k\chi}.
\end{equation}
If both $N_{\epsilon}$ and $M_{\epsilon}$ are large enough, we may further majorate the sum by the integral of the decreasing function:
\begin{multline}
    \alpha \omega_c^2\sum\limits_{k=-\infty}^{-N_{\epsilon}}\chi e^{-e^{k\chi}\tau^{\prime}\omega_c+2k\chi}<\\<\alpha \omega_c^2\int\limits^{\infty}_{N_{\epsilon}\chi} e^{-e^{-x}\tau^{\prime}\omega_c-2x}dx\overset{\text{def}}{=}\Delta_N(t),
\end{multline}
\begin{multline}
\alpha \omega_c^2\sum\limits^{\infty}_{k=M_{\epsilon}}\chi e^{-e^{k\chi}\tau^{\prime}\omega_c+2k\chi}<\\<\alpha \omega_c^2\int\limits^{\infty}_{M_{\epsilon}\chi}e^{-e^{x}\tau^{\prime}\omega_c+2x}dx\overset{\text{def}}{=}\Delta_M(t).
\end{multline}
In order to estimate the $L_1$ norm of the error, let us first integrate over $t$:
\begin{equation}
    \int\limits_{t=0}^T \Delta_N(t)\le \alpha \omega_c\int\limits_{N_{\epsilon}\chi}^{\infty}e^{-x}e^{-\frac{1}{\sqrt{2}}e^{-x}}\le
   \alpha \omega_c e^{-N_{\epsilon}\chi}
\end{equation}
\begin{equation}
     \int\limits_{t=0}^T \Delta_M(t)\le \alpha \omega_c\int\limits_{M_{\epsilon}\chi}^{\infty}e^{x}e^{-\frac{1}{\sqrt{2}}e^{x}}\le \alpha \omega_c e^{-\frac{1}{\sqrt{2}}e^{M_{\epsilon}\chi}}
\end{equation}
The estimates for $N_{\epsilon},M_{\epsilon}$ then follows. As we have seen, $N_{\epsilon}$ scales as $\log^2(\frac{\alpha \omega_c }{\epsilon_1})$ whereas $M_{\epsilon}$ scales slower, as $\log(\frac{\alpha \omega_c}{\epsilon_1})\log\left(\sqrt{2}\log(\frac{\alpha \omega_c}{\epsilon_1})\right)$. This is not surprising as $N_{\epsilon}-$ terms cover the long range behavior of the IF which is more important in this limit and carry most of the physics.

\section{Truncation of the boson}\label{Truncation of bosonic modes}
The aim of this section is to prove that the MPS matrices 
defined in the main text can be truncated to the subspace of $n_{\star}$ bosonic excitations, preserving the fixed error $\epsilon$ on the observables. 

We do it in two steps, first we prove that the excitations with large number of bosons are exponentially suppressed. Then we will estimate the effect of neglecting the excitations with $n>n_{\star}\sim \log(\frac{(\omega_cT)^2}{\epsilon})$.

Let us first consider the continuous limit $\Delta t \to 0$ of the quantum evolution
:
\begin{equation}
\mathcal{U}^{(i)}_{s_{i+1},\bar{s}_{i+1}|s_i,\bar{s}_i}= M_{s_i,\bar{s}_i} \otimes \left( U^{(i)}_{s_{i+1},s_{i}}\otimes(U^{(i)})^{\star}_{\bar{s}_{i+1},\bar{s}_{i}}\right).
\end{equation}
The wave function $\Psi_{\boldsymbol{n},\bar{\boldsymbol{n}}|s,\bar{s}}$ is labeled by a two integer valued vectors: $\boldsymbol{n}$,$\bar{\boldsymbol{n}}$ and also spin label $s,\bar{s}=\pm 1$. The equations of motion are written as:
\begin{multline}\label{continuous_eq}
    \dot{\Psi}_{\boldsymbol{n},\boldsymbol{\bar{n}}|s,\bar{s}}=-\Big({\sum\limits_{k=-N_{\epsilon}}^{M_{\epsilon}}}\Omega_k n_k+\Omega^{\star}_k\bar{n}_k\Big){\Psi}_{\boldsymbol{n},\boldsymbol{\bar{n}}|s,\bar{s}}-\\-{\sum\limits_{k=-N_{\epsilon}}^{M_{\epsilon}}}\Big[\sqrt{2} \lambda_k s\sqrt{n_k}\Psi_{\boldsymbol{n}-1_k,\bar{\boldsymbol{n}}|s,\bar{s}}-\sqrt{2}\lambda_k\bar{s}\sqrt{{\bar{n}_k}}\Psi_{\boldsymbol{n},\bar{\boldsymbol{n}}-1_k|s,\bar{s}}\Big]+\\+{\sum\limits_{k=-N_{\epsilon}}^{M_{\epsilon}}}
\lambda_k\frac{s-\bar{s}}{\sqrt{2}}\Big[\sqrt{n_k+1} \Psi_{\boldsymbol{n}+1_k,\bar{\boldsymbol{n}}|s,\bar{s}}-\sqrt{\bar{n}_k+1}\Psi_{\boldsymbol{n},\bar{\boldsymbol{n}}+1_k|s,\bar{s}}\Big]+\\+
i\sum\limits_{s'=\pm 1} h^{(i)}_{s,s'}{\Psi}_{\boldsymbol{n},\boldsymbol{\bar{n}}|s',\bar{s}}-i\sum\limits_{\bar{s}'=\pm 1} h^{(i)}_{\bar{s},\bar{s}'}{\Psi}_{\boldsymbol{n},\boldsymbol{\bar{n}}|s,\bar{s}'}.
\end{multline}
Here $h^{(i)}$ is a logarithm of a unitary $U^{(i)}=e^{ih^{(i)}\Delta t}$, acting on the spin. We note that the above equation is almost identical to an equation of motion appearing in the HEOM approach, see Ref.~\cite{xu2022taming}. This underlines the connection between the two approaches.

We will use these equations to estimate the norm squared of a wave function 
\begin{equation}
|\Psi_{\boldsymbol{n},\boldsymbol{\bar{n}}}|^2\overset{\text{def}}{=}\sum\limits_{s,\bar{s}=\pm 1} |\Psi_{\boldsymbol{n},\boldsymbol{\bar{n}}|s,\bar{s}}|^2, 
\end{equation}
we have:
\begin{multline}\label{Norm estimate}
\frac{d}{dt}|\Psi_{\boldsymbol{n},\boldsymbol{\bar{n}}}|<-\Big({\sum\limits_{k=-N_{\epsilon}}^{M_{\epsilon}}}\gamma_k (n_k+\bar{n}_k)\Big)|\Psi_{\boldsymbol{n},\boldsymbol{\bar{n}}}|+\\+4{\sum\limits_{k=-N_{\epsilon}}^{M_{\epsilon}}}\sqrt{2} |\lambda_k|\Big[\sqrt{n_k}|\Psi_{\boldsymbol{n}-1_k,\boldsymbol{\bar{n}}}|+\sqrt{\bar{n}_k}|\Psi_{\boldsymbol{n},\bar{\boldsymbol{n}}-1_k}|\Big]+
\\+4{\sum\limits_{k=-N_{\epsilon}}^{M_{\epsilon}}}
\sqrt{2}|\lambda_k|\Big[ \sqrt{n_k+1}|\Psi_{\boldsymbol{n}+1_k,\boldsymbol{\bar{n}}}|+\sqrt{\bar{n}_k+1}|\Psi_{\boldsymbol{n}+1_k,\boldsymbol{\bar{n}}}|\Big].
\end{multline}
We claim that the amplitudes fulfill the inequality:
\begin{equation}\label{bound}
|\Psi_{n,\bar{n}}|\le f_{\boldsymbol{n},\bar{\boldsymbol{n}}}\overset{\text{def}}{=}4
\prod\limits_{k=-N_{\epsilon}}^{M_{\epsilon}}\Big|\frac{4\sqrt{2}\kappa\lambda_k }{\gamma_k}\Big|^{n_k+\bar{n}_k}\frac{1}{\sqrt{n_k!\bar{n}_k!}},
\end{equation}
with $\kappa>1$ to be defined later. 

Indeed, the inequality holds for the initial wave function which is nonzero only for $\boldsymbol{n}=\bar{\boldsymbol{n}}=\boldsymbol{0}$. We will use an induction, assume that the inequality \eqref{bound} holds up to the moment $t$ and substitute it to the equation \eqref{Norm estimate}. Note that the function $f_{\boldsymbol{n},\bar{\boldsymbol{n}}}$ fulfill the following important relation:
\begin{equation}
f_{\boldsymbol{n}-1_k,\bar{\boldsymbol{n}}}=\frac{|\gamma_k|\sqrt{n_k!}}{4\sqrt{2}\kappa|\lambda_k|}f_{\boldsymbol{n},\bar{\boldsymbol{n}}},
\end{equation}
\begin{equation}
f_{\boldsymbol{n},\bar{\boldsymbol{n}}-1_k}=\frac{|\gamma_k|\sqrt{\bar{n}_k!}}{4\sqrt{2}\kappa|\lambda_k|}f_{\boldsymbol{n},\bar{\boldsymbol{n}}_k}.
\end{equation}
Using it, we may simplify \eqref{Norm estimate}:
\begin{multline} \label{Norm estimate 2}
\frac{d}{dt}|\Psi_{\boldsymbol{n},\boldsymbol{\bar{n}}}|<-\Big(\sum\limits_{k=-N_{\epsilon}}^{M_{\epsilon}}\gamma_k (n_k+\bar{n}_k)\Big)|\Psi_{\boldsymbol{n},\boldsymbol{\bar{n}}}|+\\+\frac{1}{\kappa}\Big({\sum\limits_{k=-N_{\epsilon}}^{M_{\epsilon}}}\gamma_k (n_k+\bar{n}_k)\Big)f_{\boldsymbol{n},\boldsymbol{\bar{n}}}+
\\+\sum\limits_{k=-N_{\epsilon}}^{M_{\epsilon}}
\frac{32 \kappa|\lambda_k|^2}{\gamma_k}\Big[ \sqrt{\frac{n_k+1}{n_k}}+\sqrt{\frac{\bar{n}_k+1}{\bar{n}_k}}\Big]f_{\boldsymbol{n},\boldsymbol{\bar{n}}}.
\end{multline}
Let us, for the moment, ignore the last term (we will see later that it is indeed suppressed). It is clear that $\Psi_{\boldsymbol{n},\bar{\boldsymbol{n}}}$ will always remain less than $f_{\boldsymbol{n},\bar{\boldsymbol{n}}}$, if it was true at the initial time $t$. Indeed, with the last term dropped, equation \eqref{Norm estimate 2} turns to:
\begin{multline}
\frac{d}{dt}|\Psi_{\boldsymbol{n},\boldsymbol{\bar{n}}}|<-\Big(\sum\limits_{k=-N_{\epsilon}}^{M_{\epsilon}}\gamma_k (n_k+\bar{n}_k)\Big)|\Psi_{\boldsymbol{n},\boldsymbol{\bar{n}}}|+\\+\frac{1}{\kappa}\Big(\sum\limits_{k=-N_{\epsilon}}^{M_{\epsilon}}\gamma_k (n_k+\bar{n}_k)\Big)f_{\boldsymbol{n},\boldsymbol{\bar{n}}}.
\end{multline}
Which means that the norm $|\Psi_{\boldsymbol{n},\boldsymbol{\bar{n}}}|$ always decrease, whenever it gets close to the $f_{\boldsymbol{n},\boldsymbol{\bar{n}}}$. It is useful to introduce the notation $\nu_k=\frac{32|\lambda_k^2|}{|\gamma_k^2|}$. The parameter $\nu_k$ controls the average excitation of $k$-th mode. It is important that the parameters $\lambda_k,\gamma_k$ 
are such, that $\nu_k$ may be bounded as: 
\begin{equation}
\label{Nu_bound} 
\nu_k<\nu_{\star}=64\alpha\chi\sim\frac{\alpha}{\log(\frac{\alpha \omega_c T}{\epsilon })}.
\end{equation} 
Thus, for a small enough error $\epsilon$, there is always exists $\kappa\sim 1+\frac{\alpha}{\log(\frac{\alpha \omega_c T}{\epsilon })}$ such that the last term in \eqref{Norm estimate 2} may be estimated to be smaller than the second term. Namely, we choose $\kappa$ to fulfill the inequality:
\begin{equation}
    \frac{1-\kappa}{\kappa}
\le \frac{32\sqrt{2} \kappa|\lambda_k|^2}{\gamma^2_k}
\end{equation}
for any $k$. This implies that the norm $|\Psi_{\boldsymbol{n},\boldsymbol{\bar{n}}}|$ obeys the equation:
\begin{multline} \label{Final estimate}
\frac{d}{dt}|\Psi_{\boldsymbol{n},\boldsymbol{\bar{n}}}|<-\Big(\sum\limits_{k=-N_{\epsilon}}^{M_{\epsilon}}\gamma_k (n_k+\bar{n}_k)\Big)|\Psi_{\boldsymbol{n},\boldsymbol{\bar{n}}}|+\\+\Big(\sum\limits_{k=-N_{\epsilon}}^{M_{\epsilon}}\gamma_k (n_k+\bar{n}_k)\Big)f_{\boldsymbol{n},\boldsymbol{\bar{n}}}.
\end{multline}
This finally proves the bound \eqref{bound}. The same reasoning works in case of discrete dynamics. The operator of quantum evolution $U_{\text{tot}}$ is the consecutive application of unitary $U^{(i)}$ acting on the spin and operator $M_{s_i,\bar{s}_i}$ acting on both bosonic and spin degrees of freedom. The first operator $U^{(i)}$, obviously doesn't change the norm $|\Psi_{\boldsymbol{n},\bar{\boldsymbol{n}}}|$, and the MPS operator $M_{s,\bar{s}}$ preserves the bound \eqref{bound}. Indeed, we have:
\begin{equation}
M_{s_i,\bar{s}_i}=e^{-\Lambda(s_i-\bar{s}_i)^2}e^{\Delta t\mathcal{M}_{s_i,\bar{s}_i}}.
\end{equation}
The first operator $e^{-\Lambda(s_i-\bar{s}_i)^2}$ is clearly a quantum channel, and doesn't increase the norm $|c_{\boldsymbol{n},\bar{\boldsymbol{n}}}|$.
The result of application of operator $e^{\mathcal{M}_{s_i,\bar{s}_i}\Delta t}$ may be then viewed as a continuous evolution from time $t$ to $t+\Delta t$, which also preserves the bound \eqref{bound} by the same argument as in the continuous case, this finishes the prove of the bound \eqref{bound}.

This bound dictates that the wave function decays with the growth of the number of bosons. The natural idea is to truncate the states of bosons with more than $n_{\star}$ excitations, for some fixed $n_{\star}(\epsilon)$. In order to prove the bound on the error, let us first massage the function $f_{\boldsymbol{n},\bar{\boldsymbol{n}}}$. Using the bound for $\nu_k$ \eqref{Nu_bound}, we can assume that for $\epsilon$ small enough, all $\nu_k$ are less than $1$. Thus we may use a very rough estimate:
\begin{equation}\label{f_estimate}
f_{\boldsymbol{n},\bar{\boldsymbol{n}}} \le 4\nu_{\star}^{\sum\limits_{k=-N_{\epsilon}}^{M_{\epsilon}}\frac{n_k+\bar{n}_k}{2}}.
\end{equation}

To estimate the corresponding error, let us split the Hilbert space in two spaces: $\mathbb{H}_l$ consisting of excitations with $n_{\star}$ bosons or less, and $\mathbb{H}_e$ with more than $n_{\star}$ bosons. By projecting out the $\mathbb{H}_e$ subspace - we exclude the processes when particle enter this subspace, spend some time $\tau=\Delta t m$ inside and then relax back to $\mathbb{H}_l$. The amplitude of this process is given by: 
\begin{equation}
A_m=\langle \Psi_l|\mathbb{P}_l \left(\mathcal{U}\ \mathbb{P}_e\right)^{m-1}\mathcal{U}\mathbb{P}_l|\Psi_r\rangle,
\end{equation}
where $\mathbb{P}_l,\mathbb{P}_e$ are the projectors to the low/excited spaces respectievely.

As we just have proved, all the intermediate states fulfill the inequality \eqref{bound}. Because of the suppression of the excited states, we conclude that each amplitude may be estimated by the transition rate:
\begin{equation}
    |A_m|\le \Gamma\overset{\text{def}}{=}\max\limits_{\Psi_l,\Psi_r}|\langle\Psi_l|\mathbb{P}_l \mathcal{U}\ \mathbb{P}_e\mathcal{U}\mathbb{P}_l|\Psi_r\rangle|.
\end{equation}
In order to estimate $\Gamma$ we again use the equation \eqref{Final estimate}. One obtains:
\begin{multline}
\Gamma\le 16\nu_{\star}^{\sum\limits_{k=-N_{\epsilon}}^{M_{\epsilon}}(n_k+\bar{n}_k)}\left(1-e^{-\sum\limits_{k=-N_{\epsilon}}^{M_{\epsilon}} \gamma_k\Delta t(n_k+\bar{n}_k)}\right)^2\le \\ 16\nu_{\star}^{n_{\star}}\left(\sum\limits_{k=-N_{\epsilon}}^{M_{\epsilon}} \gamma_k\Delta t(n_k+\bar{n}_k)\right)^2.
\end{multline}
In the last inequality we used the fact that (as we will see below) the sum $\sum\limits_{k=-N_{\epsilon}}^{M_{\epsilon}} \gamma_k (n_k+\bar{n}_k)$ scales polylogarithmically in time, while $\Delta t$ scales as $\frac{\epsilon}{T}$, and so the argument in the exponent is less than $1$.
As $\{\gamma_n\}$ is an increasing sequence
, we can also bound
\begin{equation}
\gamma_k<\frac{\omega_c}{\sqrt{2}}\log(\frac{\alpha \omega_c}{\epsilon_1}).    
\end{equation}
Which imply a bound for a $\Gamma$:
\begin{equation}
\Gamma\le 8\nu_{\star}^{n_{\star}}\left(\Delta t \omega_c\log\left(\frac{\alpha \omega_c}{\epsilon_1}\right)n_{\star}\right)^2.
\end{equation}
The total error scales as:
\begin{equation}
\epsilon=N^2 \Gamma=8\nu_{\star}^{n_{\star}}\left(T \omega_c\log\left(\frac{\alpha \omega_c}{\epsilon_1}\right)n_{\star}\right)^2,
\end{equation} 
such that we can choose
\begin{equation}\label{n_star}
n_{\star}\sim \frac{\log(\frac{\left(\omega_cT\right)^2}{\epsilon})}{\log(\nu_{\star}^{-1})}.    
\end{equation}
If we denote the total number of modes $N_{\epsilon}+M_{\epsilon}=K\sim \log^2\left(\frac{\omega_c T}{\alpha}\right)$. The the total number of states scales as a sum of binomial coefficients
\begin{equation}
    D=\sum\limits_{n=1}^{n_{\star}}C_{n+K}^K=\sum\limits_{n=1}^{n_{\star}}\frac{(n+K)!}{n!K!}.
\end{equation}
For the estimate we can assume that $K>n_{\star}$, such that:
\begin{equation}
    \frac{(n+K)!}{K!}<(2K)^{n}<(2K)^{n_{\star}}.
\end{equation}
Putting all together, we obtain: 
\begin{equation}\label{D_final}
D<e(2K)^{n_{\star}}\sim e^{\# \frac{\log\left(\frac{(\omega_cT)^2}{\epsilon}\right)}{\log(\nu_{\star}^{-1})}\log(\log\left(\frac{\alpha \omega_cT}{\epsilon}\right))}.
\end{equation}
Asymptotically, when we increase the accuracy $\epsilon\to 0$, for fixed coupling strength $\alpha$, the fraction $\frac{\log(\log\left(\frac{\alpha \omega_c T}{\epsilon}\right))}{\log(\nu_{\star}^{-1})}$ limits to one. And we arrive to a simple estimate
:
\begin{equation}
D\to\frac{(\omega_cT)^4}{\epsilon^2}
\end{equation}
Which is the main statement of the current paper.
\section{From quantum channel to an auxiliary zero temperature boson}
The way we introduced bosons in the main text was rather technical. In this short section we provide a connection with a physical picture. Namely, for the positive $\lambda$, the very same IF generated by MPS:
\begin{equation}
I_{\{\boldsymbol{s},\boldsymbol{\bar{s}}\}}=\langle  0 |M_{s_N,\bar{s}_N}\dots M_{s_1,\bar{s}_1}|\rho_b\rangle.
\end{equation}
May be generated by a quantum channel:
\begin{multline}
    \mathcal{E}(\rho_b\otimes |s_i\rangle \langle \bar{s}_i|)=\sum\limits_{n=0}^{\infty}E_n(\rho_b\otimes |s_i\rangle \langle \bar{s}_i|)E^{\dagger}_n=\\=\sum\limits_{n=0}^{\infty}\Big[\frac{\kappa^n}{n!} a^n e^{-\Omega \Delta t a^{\dagger}a}e^{-\lambda\Delta t s_i a^{\dagger}}e^{\lambda \Delta t s_i a}\left(\rho_b\otimes |s_i\rangle \langle \bar{s}_i|\right)\times \\ \times e^{-\lambda\Delta t \bar{s}_i a^{\dagger}} e^{-\lambda\Delta t \bar{s}_i a}e^{-\Omega^{\star}\Delta t a^{\dagger}a}\left(a^{\dagger}\right)^n\Big].
\end{multline}
The condition for $\mathcal{E}$ to be a quantum channel is
\begin{equation}
\sum\limits_{n=0}^{\infty}E_n^{\dagger}E_n=1,  
\end{equation}
which is equivalent to $\kappa=(e^{2\gamma \Delta t}-1)$.

In order to see the equivalence, we introduce the density matrix/state duality: $\rho_b\to  \sum\limits_{n=0}^{\infty}\rho_b|n\rangle\otimes |n\rangle$, we then may rewrite the quantum channel as:
\begin{multline}
\mathcal{E}(\rho_b\otimes |s_i\rangle \langle \bar{s}_i|)=\mathcal{E}_{s_i,\bar{s}_i}|\rho_b\rangle |s_i\rangle|\bar{s}_i\rangle =\\=e^{\kappa a \bar{a}}e^{-\Omega\Delta t a^{\dagger}a-\Omega^{\star}\Delta t\bar{a}^{\dagger}\bar{a}}\times \\ \times e^{-\lambda\Delta t s_ia^{\dagger}-\lambda\Delta t \bar{s}_i\bar{a}^{\dagger}}e^{s_i\lambda \Delta ta+\bar{s}_i\lambda\Delta t\bar{a}}|\rho_b\rangle |s_i\rangle|\bar{s}_i\rangle
\end{multline}
Direct calculation shows that the multiple application of quantum channel $\mathcal{E}_{s_i,\bar{s}_i}$ and trace over bosonic space is equivalent to the vacuum expectation value of multiple application of ZMKMP operator $M_{s_i,\bar{s}_i}$.
\begin{multline}\label{qChannel_to_ZMKMP}
I_{\{\boldsymbol{s},\boldsymbol{\bar{s}}\}}=\langle  0 |e^{a \bar{a}}\mathcal{E}_{s_N,\bar{s}_N}\dots\mathcal{E}_{s_1,\bar{s}_1}|\rho_b\rangle=\\=\langle  0 |M_{s_N,\bar{s}_N}\dots M_{s_1,\bar{s}_1}|\rho_b\rangle.
\end{multline}

\end{document}